\documentclass[12pt]{iopart}
\usepackage{iopams}

\def\real{{\rm I\kern-.2em R}}
\def\complex{\kern.1em{\raise.47ex\hbox{  $\scriptscriptstyle
|$}}\kern-.40em{\rm C}}
\def\Box{\hfill\vbox{\hrule height 0.6pt
        \hbox{\vrule width 0.6pt height 1.8ex \kern 1.8ex
                \vrule width 0.6pt}
   \hrule height 0.6pt}}

\newtheorem{thm}{Theorem}
\newtheorem{thm1}[thm]{Theorem}
\newtheorem{lemma}{Lemma}
\newtheorem{lemma1}[lemma]{Lemma}

\newtheorem{proposition}{Proposition}
\newtheorem{proposition1}[proposition]{Proposition}

\newtheorem{lemma4}[lemma]{Lemma}
\begin{document}

\title{Smooth adiabatic evolutions with leaky power tails}
\author{J E Avron and A Elgart
\footnote[1]{E-mail addresses: {\tt avron@physics.technion.ac.il}
and {\tt elgart@physics.technion.ac.il}}}

\address{Department of Physics, Technion, 32000 Haifa, Israel.}
\begin{flushright}
\it\small {Tosio Kato--in memoriam}
\end{flushright}

\begin{abstract}
Adiabatic evolutions {\em with} a gap condition have, under a
range of circumstances, exponentially small tails that describe
the leaking out of the spectral subspace. Adiabatic evolutions
{\em without} a gap condition do not seem to have this feature in
general. This is a known fact for eigenvalue crossing. We show
that this is also the case for eigenvalues at the threshold of the
continuous spectrum by considering the Friedrichs model.
\end{abstract}
\section{Introduction}
Adiabatic theorems describe the solutions of  initial value
problems where the Hamil\-tonian generating the evolution depends
slowly on time.  In quantum mechanics the description is in terms
of  spectral information of the instantaneous Hamiltonian. A few
basic references on various types of
  adiabatic theorems
 are \cite{b,gp,j,kato2,mn,n}.

 To formulate the problem  more
precisely it is convenient to replace the physical time $t$ by the
scaled time $s=t/\tau$. One is then concerned with the solution of
the initial value problem
\begin{equation} i\, \dot \psi_\tau (s) = \tau H(s)\, \psi_\tau
(s),\label{schrod}
\end{equation}
 with initial data
$\psi_{\tau}(0)\in Range P(0)$. $P(s)$ is a spectral projection for $H(s)$,
 a self-adjoint
Hamiltonian which depends sufficiently smoothly on $s$.
$\psi_{\tau}$ is a vector valued function and the adiabatic limit is the limit
of large $\tau$. Suppose, for the sake
of concreteness, that $\dot H(s)$ is compactly supported on
$[0,1]$.
  As $s$ runs on the interval $[0,1]$, $H(s)$ evolves slowly in
 {\em physical} time for a {\em long} interval
 of time.
 The total   variation of  $H(s)$ is finite and not
 necessarily small.

 Adiabatic theorems fall into two baskets: Those
that describe the solutions for {\em all} times, including times
 $s\in [0,1]$, and those that characterize the solutions
at large times $s>1$ where the Hamiltonian is time independent
again. Interestingly they give  more precision for long times. We
call the first basket, the one that applies to all times, {\em
uniform}, the second is the {\em long time} basket.

Adiabatic theorems can also be put in two other baskets. Those
that satisfy a gap conditions and those that do not. By a gap
condition we mean a spectral condition on the Hamiltonian $H(s)$
so that a finite gap separates the spectral subspace of the
initial data ($=Range P(s)$), from the rest of the spectrum for
all times $s$.

A representative result from the uniform basket is \cite{ae1,b}:
\begin{thm}\label{uniform}
Suppose that $P(s)$ is smooth finite rank spectral projection, for
a bounded, smooth Hamiltonian $H(s)$. Then, the evolution of the
initial state $\psi_{\tau}(0)\in Range P(0)$,  is such that
\begin{equation} dist\Big(\psi_\tau (s), Range P(s)\Big) \le o(1)
\end{equation} for all
$s\ge 0$. If, in addition,  $P(s)$ satisfies a gap condition then
\begin{equation} dist\Big(\psi_\tau (s), Range P(s)\Big) \le O(1/\tau)
\end{equation} for all
$s\ge 0$.
\end{thm}
{\it {Remark 1}}\,\,\, By $o(1)$ we mean a term that vanishes as
$\tau\to\infty$.
\newline
{\it {Remark 2}}\,\,\, Schr\"odinger operators is a class of
application not covered by the theorem because of the assumption
that $H(s)$ is bounded. This restriction can be lifted by standard
machinery see e.g. \cite{ae1,b}. We chose not to enter into this
because the essence of the adiabatic evolution is an infrared
problem which is largely divorced from the issue of unboundedness
of the generator (an ultra-violet problem).

A characteristic result which lies both in the long time basket and in the
gap condition basket is
 \cite{berry,ks,g,martinez,N,js,joye}:
\begin{thm1}\label{th2}
Suppose that $H(s)$ is a smooth, bounded, and self-adjoint with
$\dot H(s)$ supported on $[0,1]$. And suppose that $P(s)$ is
separated by a  finite gap from the rest of the spectrum. Then, the
evolution of the initial state $\psi_{\tau}(0)\in Range P(0)$, is
such that for any $s>1$,
\begin{equation}dist \Big(\psi_\tau(s), Range P(s)\Big) =
o\left(\frac{1}{\tau^n}\right)\end{equation} for all $n\ge 0$.
\end{thm1}
{\it {Remarks }}
\begin{enumerate}
\item There is, in general, no uniformity in $n$;
 the  term on the right hand
side is of order $\frac{c_n}{\tau^n}$ where $c_n$ are allowed to
grow rapidly with $n$.
\item
 In the case where
$\dot H(s)$ is an analytic family which decays at infinity,
the leaking at $s=\infty$ is exponentially small.
\item In the case of a gap  the
distinction between the uniform and the long time basket has an
analog in integrals. Suppose that $g(s)$ is real, smooth, strictly
monotonic and compactly supported in $[0,1]$. Then
\[\int_{-\infty}^s g(t) e^{it\tau} dt =\left\{
\begin{array}{ll}
o(1/\tau^n)& \mbox{if $s>1$, for all $n$;} \\ O(1/\tau)& \mbox{if
$s\in[0,1]$.}
\end{array}
\right.
\]The gapless basket seems to have no analog in integrals, as we
shall see.
\end{enumerate}

What can one say about adiabatic theorems that lie  both in the
long time and gapless baskets? In the uniform adiabatic
 theorem \ref{uniform},
the price one pays for the absence of a gap is, in general,
loss of information on the rate at which the adiabatic limit is
approached.
The question we address here is what is the price one pays in
the long time adiabatic theorem, for the absence of a gap.

Adiabatic theorems without a gap condition are normally associated with
two distinct settings:
 Those that  describe  eigenvalue crossings and those associated
with eigenvalue at the threshold of the essential spectrum.

Hagedorn \cite{hagedorn} studied adiabatic theorems for  crossing
eigenvalues. He showed that
 {\em there is no improvement in the adiabatic theorem in
long time limit.} In particular, the leaking is always of order
$\frac{1}{\sqrt \tau}$ for linear crossing.

Here we consider the cases where the absence of a gap is
associated with an eigenvalue at the threshold of the continuous
spectrum. The results of Hagedorn for crossing eigenvalues do not
shed light on this case\footnote{At crossing the spectral
projection is discontinuous while for the eigenvalue at threshold
that we consider the spectral projection is smooth for all
times.}.

To investigate this we consider  the Friedrichs model
\cite{friedrichs1} with an eigenvalue at the threshold of the
continuous spectrum. The model is parameterized by a real
parameter $\beta>0$ related to the behavior of the density of
states at low energies. Our main result is\footnote{All estimates
 should be considered in asymptotic sense with respect to
$\tau\rightarrow\infty$.}
\begin{equation}dist\Big(\psi_\tau
(s), Range P(s)\Big) =\left\{\begin{array}{ll}
O(\tau^{-\beta})&\mbox{$1<\beta<2$ and $s>1$;}\\
O(\tau^{-1})&\mbox{$1<\beta<2$ and $s\in[0,1].$}
\end{array}\right.\end{equation}
Unlike the gap basket, the long time behavior in the Friedrichs
model indeed has power tails, and unlike the crossing basket,
there is an improvement in the rate of decay at long times.

The main application of adiabatic theorems with eigenvalues at
threshold is to the theory of atoms in radiation field, where the
photon field eliminates spectral gaps. The absence of a gap makes
the adiabatic theory of atoms in radiation field qualitatively
different from quantum mechanics. Two distinct processes lead to
the error in adiabatic evolutions. The first is photon production
and the second is atomic excitation. Folk wisdom is that the
leaking due to atomic excitations has exponential tails. A simple
argument suggests that the leaking due to photon production is of
order $1/\tau$. The argument goes as follows: The power radiated
by a dipole is, by classical electrodynamics, $$\frac{2}{3}
\frac{(\dot d)^2}{c^3},$$ where $d$ is the dipole moment and $c$
the velocity of light. It follows that the number of radiated
photons (assuming characteristic frequency of order
$O(1/\tau)$\,), is  of order $$\left(
\frac{\alpha^{3/2}}{\tau}\right)^2,$$ in atomic
units\footnote{Recall that in atomic units $\alpha=1/c$.}.

\section {Adiabatic evolutions}
A basic strategy for  studying adiabatic
evolutions, introduced by
Kato \cite{kato2}, is to compare the true evolution, $U_\tau(s)$  with a
fictitious one, $U_a(s)$, which respects the spectral
splitting. That is
\begin{equation}\label{ad-ev}
P(s)=U_a(s)PU^\dagger_a(s).
\end{equation}
 We shall denote by a subscript  the generator  and the evolution
 of such a fictitious
dynamics, i.e.
\begin{equation}
i\dot U_a(s)=\tau H_a(s)U_a(s),\,  U_a(0)=1,
\end{equation}
while for the true dynamics:
\begin{equation}
i\dot U_\tau(s)=\tau H(s)U_\tau(s),\,  U_\tau(0)=1.
\end{equation}
It was shown by Kato that $H_a$
satisfies the commutator equation:
\begin{equation}
\tau [H_a(s),P(s)] = i\,[\dot P(s),P(s)].
\end{equation}
This commutator equation does not have a unique solution, and
different choices of $H_a$ can be made.  Kato chose:
\begin{equation}
H_{K}=\frac{i}{\tau}[\dot P(s),P(s)]\,.
\end{equation}
This generates a geometric evolution so that, in particular,
$U_K(s)$ is
 independent of $\tau$. It  is a convenient choice for proving
  a  uniform adiabatic theorem when $P(s)$ is the kernel of
 $H(s)$. It is inappropriate for generating a systematic adiabatic
 expansion.

A more effective choice, introduced in \cite{asy}, is:
\begin{equation}
H_{AD}=H(s)+\frac{i}{\tau}[\dot P(s),P(s)]\,.
\end{equation}
This generator is  close to the Hamiltonian and  it can be
used to prove stronger adiabatic theorems than those handled by Kato,
 and it can also be used to generate an adiabatic expansion and
prove  long time adiabatic theorems \cite{ks}. However, it is
often difficult to use this evolution for concrete computations
because there is no explicit formula for
 $U_{AD}(s;\tau)$.
This is what one needs to do when one wants to show fat leaky
tails where one needs a {\em lower bound} on the part that leaks.
For this reason we shall need  yet another choice of an adiabatic
evolution that more readily lends itself to explicit computation.

In the special case when
the time dependence enters through the unitary family
\begin{equation}
H(s)=V(s)HV^\dagger(s)\, ,
\end{equation}
the evolution in the rotating frame is both adiabatic and explicit:
\begin{equation}
U_{r}(s)=V(s)\exp{(-i\tau s H)}\, .
\end{equation}
It is generated by
\begin{eqnarray}
H_{r}(s)&=&H(s)+\frac{i}{\tau}\dot V(s)V^\dagger(s)\,
,\end{eqnarray}
and like $H_{AD}$ is also close
to $H(s)$. This evolution turns out to be inferior to $H_{AD}$
 when the game is to prove
{\em upper bounds} on the leaky tails, but, it is a useful
evolution in estimating fat leaky tails where the game is to get
lower bounds. The best of all worlds is when $H_{AD}$ and $H_r$
coincide. This will turn out to be the case in the Friedrichs
model we consider and this is what makes the analysis of this case
simple.
\section{The Wave Operator}
To compare the true dynamics with the fictitious dynamics one
introduces the wave operator
\begin{equation}
\Omega_{AD}(s;\tau)=U^\dagger_{AD}(s;\tau)U_\tau(s).
\end{equation}
The leaking out of the spectral subspace is governed by the ``off
diagonal'' part of $\Omega_{AD}$. Namely, from estimate on
$P\Omega_{AD}(s;\tau)P_\perp$ and $P_\perp\Omega_{AD}(s;\tau)P$
where $P=P(0)$ and $P_\perp=1-P$. This follows from
\begin{eqnarray}
\Vert P_\tau(s)-P(s)\Vert&=& \Vert\,P_\perp\Omega_{AD}(s;\tau) P
-P\Omega_{AD}(s;\tau) P_\perp\,\Vert \nonumber\\&=&
\max\big\{\Vert P_\perp\Omega_{AD}(s;\tau) P\Vert\,,\,\Vert
P\Omega_{AD}(s;\tau)P_\perp\Vert\big\},\label{offdiag1}
\end{eqnarray}
 where
\begin{equation}P_\tau(s)= U_\tau(s)
PU^\dagger_\tau(s)\, .\end{equation}

$\Omega_{AD}$ can be calculated via a Volterra-type equation:
\begin{equation}\label{Volt}
\dot \Omega_{AD}(s;\tau)=K_{AD}(s)\Omega_{AD}(s;\tau)\,,
\end{equation}
where \begin{equation}K_{AD}(s) =-U^\dagger_{AD}(s;\tau)\,[\dot
P,P](s)\,U_{AD}(s;\tau).\end{equation}

By standard arguments, the series
\begin{equation}\label{expand}
\Omega_{AD}(s;\tau)=\sum_0^\infty\Omega_i(s)\, ,\quad
\Omega_0(s)=1,\quad \Omega_{i+1}(s)=\int_{0}^s K_{AD}(t)
\Omega_i(t)dt\end{equation} is absolutely
convergent\footnote{Recall that we assume $H(s), $ and $\dot P(s)$
bounded.}.

It is not {\it a-priori} clear that the series  in
Eq~(\ref{expand}) is an expansion in the small parameter of the
adiabatic limit, $1/\tau$. To see that it does note:
\begin{lemma}\label{main-lemma}
     \ \ \ \
\begin{enumerate}
\item $\Omega_{2j}$ is diagonal in the sense that it maps $
Range\, P$ to
 $ Range\, P$ and $ Range\, P_\perp$ to $Range\, P_\perp$, at the same
 time
 $\Omega_{2j+1}$ is off diagonal in the sense that it maps $ Range\, P$ to
 $ Range\, P_\perp$ and $ Range\, P_\perp$ to $Range\, P$.
In particular, only the odd terms contribute to leaking.
\item
 Let us denote  $Q(s):=1-\Omega_{AD}(s),\,f(\tau):= \sup_s{\Vert
Q(s)\Vert}$. Then for all $s$,
\begin{equation}
 \Vert\Omega_{i+2}(s)\Vert<
 C\,f(\tau)\sup_s{\Vert\Omega_i(s)\Vert}\,.
\end{equation}
In particular
\begin{eqnarray}
&& \Vert\Omega_{j}(s)\Vert<C\,
 f(\tau), \quad j= 1,\, 2;\nonumber \\
&& \Vert\Omega_{j}(s)\Vert<C\,
 f^2(\tau), \quad j\ge 3.
\end{eqnarray}
\end{enumerate}
\end{lemma}
{\it {Remarks}}
\begin{enumerate}
\item In the case where a gap condition holds, the
uniform adiabatic theorem says that $f(\tau)= O(1/\tau)$. It
follows that  ${\rm sup}_s{\Vert \Omega_{2i+1}(s)\Vert}\le
\frac{C}{\tau^{i+1}}$. In the absence of a gap condition $f(\tau)$
may, in general, decay more slowly with $\tau$.
\item For the lemma to be useful one needs
a strong version of the uniform adiabatic theorem which guarantees
that $Q(s)$ is small for large $\tau$. This goes beyond the
information given by the uniform adiabatic theorem quoted in the
previous section. That is, the adiabatic evolution must
approximate the true evolution both on $Range \, P$ and on its
complement $Range \,P_\perp$.
\item Note that from the estimates above it follows, that when
 $\Omega_1(s)>>f^2(\tau)$ for $s>1$ then
the leaky tails are determined by $\Omega_1(s)$.
\end{enumerate}
 {\it {Proof}}\,\,\,
The first part of the lemma is standard \cite{asy} and follows
from the
fact that $[\dot P,P]$ if off diagonal.
 The second part follows from the identity:
\begin{eqnarray}
\Omega_{i+2}(s) &=& \int_{0}^s Q(t)K_{AD}(t) \Omega_i(t)d
t-\int_{0}^s K_{AD}(t)
 \Omega_{AD}(t)Q^\dagger(t)
\Omega_{i+1}(t)d t \nonumber\\ &-& Q(s)\Omega_{i+1}(s) \,.
\end{eqnarray}
$K_{AD}(s)$ vanishes outside the interval $[0,1]$, hence,
\begin{eqnarray}
{\Vert\Omega_{i+2}(s)\Vert} &\le&{f(\tau)} \sup_s\Big( \Vert
K_{AD}(s)\Vert\cdot\Vert\Omega_i(s)\Vert + (\Vert
K_{AD}(s)\Vert+1)\Vert\Omega_{i+1}(s)\Vert\Big)\nonumber\\ &<& 2
{f(\tau)} (\sup_s{\Vert
K_{AD}(s)\Vert}+1)\sup_s{\Vert\Omega_i(s)\Vert} \,.\label{offdiag}
\end{eqnarray}
\Box
\section {Power Tails in Friedrichs Models}
The  Friedrichs model\footnote{In some circles this is known as
the Fanno model.} is defined on the Hilbert space ${\cal H}=
\complex\oplus L^2(\real_+,d\mu(k))$ with $inf(support\, \mu)\geq
0$ and $ \mu (0)=0$. A vector $\Psi\in{\cal H}$ is normalized by
\begin{equation}
\Psi=\left(\begin{array}{c} \omega\\ \psi(k)
\end{array}\right), \quad \Vert \Psi\Vert^2= |\omega\vert^2
+\int_{\real_+} \vert \psi(k)\vert^2 d\mu(k),\quad
\omega\in\complex.
\end{equation}
The Friedrichs  Hamiltonian $H$ acts
 on ${{\cal{H}}}$ like:
\begin{equation} H\,\Psi=\left(\begin{array}{ll} 0&0\\ 0&k
\end{array}\right)\,\left(\begin{array}{c}
\omega\\ |\psi\rangle
\end{array}\right)=\left(\begin{array}{c} 0\\ |k\,\psi\rangle
\end{array}\right).
\end{equation}
It has a ground state at threshold and the projection on the ground state
 $P$ is
\begin{equation} P\, =\left(\begin{array}{ll} 1&0\\ 0&0
\end{array}\right)\,.
\end{equation}
 We consider the case when the time dependence is of the form
 $H(s)=V(s) H V^\dagger(s)$ with unitary $V(s)$ generated by
\begin{equation}\label{miracle} \dot V(s) V^{\dagger}(s)=[\dot P(s),P(s)]=
i\dot g(s)\left(\begin{array}{ll} 0 & \langle \varphi|\\
|\varphi\rangle&0
\end{array}\right)\,,
\end{equation}
where $0\le\dot g(s)\in C^\infty_0([0,1])$, and
\begin{equation}
\int_0^x| \varphi|^{2}\, d\mu (k) = O(x^{2\beta}),  \quad \beta\ge
0 ,\label{estim}
\end{equation}
for small $x$. Note that due to the particular form of time
dependence (Eq.~(\ref{miracle})) one has that $H_{AD}=H_r$.

We now borrow a results from \cite{ae1}:
\begin{proposition}\label{friedrichs}
For the Friedrichs model and the adiabatic evolution
 generated by $H_{AD}$,
$f(\tau)$ of lemma \ref{main-lemma} is such that
\begin{equation}
f(\tau) \le
\left\{\begin{array}{lr}O\left(\frac{1}{\tau}\right),&\beta
>1;\\
 O\left(\frac{\log \tau}{\tau}\right),&\beta =1;\\
O\left(\tau^{-\beta}\right),&\beta < 1,
\end{array}\right.\label{uat}
\end{equation}
 for all $s$.
\end{proposition}
Now, we come to the main result of this section:
\begin{proposition1}\label{friedrichs1}
For the Friedrichs model the evolution of the state that starts as
the bound state $\psi_{\tau}(0)\in Range P$, is such that for any
$\epsilon>0$
\begin{equation}
C_1\tau^{-\beta+\epsilon}>dist\Big(\psi_\tau (s), Range P(s)\Big)>
C_2\tau^{-\beta}\,,
\end{equation}
when $2>\beta>0$ and $s>1$.
\end{proposition1}
 {\it {Proof}}\,\,\,
With the choice of $H_{AD }$ all the even terms $\Omega_{2j}$ are
diagonal while all the odd terms are off diagonal.
 By Lemma (\ref{friedrichs}), $\Omega_j= O\big(f^2(\tau)\big)$ for $j\ge 3$.
We need to compute  $\Omega_1(s)$  and provided it dominates $f^2(\tau)$,
we are done.
Now, for $s>1$,
\begin{eqnarray}
&&P_\perp \Omega_1(s)P=P_\perp \int_{0}^1 K(t) d s P\nonumber\\
&=&i \left(\begin{array}{ll} 0 &0\\ 0&1
\end{array}\right)\int_{0}^1 \dot g(s)\exp{(i\tau s
\left(\begin{array}{ll} 0 &0\\ 0&k
\end{array}\right))}\left(\begin{array}{ll} 0 & \langle
 \varphi|\\ |\varphi\rangle & 0
\end{array}\right) \left(\begin{array}{ll} 1 & 0\\ 0 & 0
\end{array}\right) d s
\nonumber\\ &=& i\int_{0}^1\dot g(s) \left(\begin{array}{ll} 0
&0\\ |\exp{(i\tau s k)}\varphi\rangle&0
\end{array}\right)d s\nonumber\\ &=&i\left(\begin{array}{ll}
0 &0\\ |\widehat{\dot g}(\tau k)\varphi\rangle&0
\end{array}\right)\,,
\end{eqnarray}
where $\widehat{\dot g}$ stands for Fourier transform of $\dot g$.
Now, since $\dot g$ is positive $|\hat{\dot g}(k)|$ takes its
maximal value at the origin. Since this function is continuous,
for some $a,\,b$ positive
 $|\widehat{\dot g}(k)|>b$ for $k\in[0,a]$. We can
estimate now the norm of $P_\perp \Omega_1(s)P$ from below using
Eq.~(\ref{estim}):
\begin{eqnarray}
\Vert P_\perp \Omega_1(s)P\Vert^2&=& \Vert  |\widehat{\dot g}(\tau
k)\varphi(k)\rangle\Vert^2 = \int_0^\infty |\widehat{\dot g}(\tau
k)|^2|\varphi(k)|^2d\mu(k)\nonumber \\ &>&\int_0^a|\widehat{\dot
g}( k)|^2|\varphi(\frac{k}{\tau})|^2d\mu(\frac{k}{\tau})\nonumber
\\ &>& b^2\int_0^\frac{a}{\tau}|\varphi(k)|^2d\mu(k)\nonumber \\
&=&O\Big(\frac{1}{\tau^{2\beta}}\Big) ,\quad s>1.
\end{eqnarray}
 Similarly,
\begin{eqnarray}
\Vert P_\perp \Omega_1(s)P\Vert^2&=& \int_0^\infty |\widehat{\dot
g}(\tau k)|^2|\varphi(k)|^2d\mu(k)\nonumber \\
&=&\{\int_0^\frac{1}{\tau^{1-\epsilon}}+\int_\frac{1}{\tau^
{1-\epsilon}}^\infty\}|\widehat{\dot
g}(\tau k)|^2|\varphi(k)|^2d\mu(k) \nonumber
\\ &<&C\Big(\frac{1}{\tau^{2\beta(1-\epsilon)}}+ |\widehat{\dot
g}(\tau^\epsilon)|^2\Big),\quad s>1.
\end{eqnarray}
Since $|\hat{\dot g}(k)|$ is decaying faster then any power at
large $k$, the proof is complete. \Box
\newline Note that there is
 an improvement in the long time limit over the uniform result
 for $2>\beta\geq 1$.

\textbf{Acknowledgment} We thank M.V. Berry for several useful
conversations. This research was supported in part by the Israel
Science Foundation, the Fund for Promotion of Research at the
Technion and the DFG.

\begin{appendix}

\section {Slaved Leaking}
In this appendix we want to address an
 apparent puzzle associated with the
 (standard) long time adiabatic theorem.

It is, in fact, surprising that a gap condition is all one needs
for the fast decay in the long time adiabatic theorem: Suppose
that $H(s)$ had one eigenvalue separated by a gap from the rest of
the spectrum, which is purely absolutely
 continuous. For time $s\in [0,1]$ the leaking to the (instantaneous)
 absolutely continuous spectrum is $O(1/\tau)$. One would expect
 that once a piece leaks to the absolutely continuous part of the
 spectrum it would propagate to infinity, as states in the absolutely
 continuous
 spectrum invariably
 do. If this was the case, then it would be difficult to understand
 how the wave reconstructs itself so that, for
 times $s>1$, the part in the absolutely
 continuous spectrum
 is smaller than any power in $1/\tau$.
As we shall see from the proof of the adiabatic theorem to all orders
below,
 the parts that leaks to the absolutely
 continuous spectrum and is a finite power of
 $\tau$ is
  slaved to the instantaneous eigenvalue, and disappears for time
  $s>1$. It does not propagate. Only
 the terms smaller than any power in $1/\tau$ are free to propagate to
 infinity.

The following lemma is taken, verbatim, from \cite{richter}
\begin{lemma1}
Let $R(s,z)$ be the resolvent of $H(s)$ and define the tilde operation by
\begin{equation}\label{twiddle}
\tilde X(s) = -\frac{1}{2\pi i} \int_\Gamma\, R(s,z)X(s)R(s,z)\, dz
\end{equation}
where $\Gamma$ is a contour enclosing the part of the spectrum associated
 with $P(s)$.  Let, $X(s) $ and $Y(s)$ be bounded, then
\begin{eqnarray}\label{main-lemma2}
P_\perp \, \int_0^s U^\dagger_{AD}(t)X(t) U_{AD} P Y(t)dt &=&
\frac{i}{\tau}\, P_\perp\left( \phantom{\int_0^0}
U^\dagger_{AD}(t)\tilde X(t) U_{AD} P Y(t )\vert_0^s\right.\nonumber \\ -
\int_0^s U^\dagger_{AD}(t)\dot{\tilde X}(t) U_{AD} P Y(t)dt&-&\left.
\int_0^s U^\dagger_{AD}(t)\tilde X(t) U_{AD} P \dot Y(t)dt\right)
\end{eqnarray}
\end{lemma1}
If we now apply the lemma with $X(s)=[\dot P,P](s)$ to $\Omega_{j+1}$ we get
\begin{eqnarray}
P_\perp\Omega_{j+1}(s)&=&P_\perp \, \int_0^s U^\dagger_{AD}(t)
X(t) U_{AD} P \Omega_j(t)dt  \\ &=&-\frac{i}{\tau}\, P_\perp\left(
 U^\dagger_{AD}(s)\tilde X(s)U_{AD}(s) P \Omega_j(s)\phantom{\int}\right.
\nonumber \\ &-&\int_0^s U^\dagger_{AD}(t)\dot{\tilde{ X}}(t)
U_{AD} P \Omega_j(t)dt\nonumber \\ &-&\left. \int_0^s
U^\dagger_{AD}(t)\tilde X(t) U_{AD} P
K(t)\Omega_{j-1}(t)dt\right)\nonumber
\end{eqnarray}
Each of the {\em integrals} in the above expression is once again
of the form so that the lemma can be applied again and again. We
see that the power terms that occur for $0<s<1$ are boundary
terms, proportional to $\dot P(s)$, and
 instantaneously reflect what happens to the bound state. That
is, the power tails  in $Range\, P_\perp(s)$ are
 slaved to the state in $Range \, P(s)$.

\section{Asymptotics of Fourier Transforms in $C_0^\infty$}
When $g\in C^\infty_0$ its Fourier transform decays faster than any power,
 but not quite exponentially. The canonical example of such a function is
\begin{equation}
g(s)=\exp\left(\frac{-1}{(1-s)(1+s)}\right),\quad s\in[-1,1]
\end{equation}
and zero otherwise. It is of some interest to have an explicit
asymptotic expansion of its Fourier transform.
\begin{lemma4}
$\widehat g(p)$ is an entire even function of $p$, and on the real $p$-axis
it has the asymptotic behavior:
\begin{eqnarray}
\widehat{g}(p) =
2\,\frac{\sqrt\pi}{(2e)^{1/4}}\,\cdot\frac{\exp(-\sqrt
p)}{p^{3/4}}\cdot\left(\cos(p-\sqrt p -\frac{3}{8}\pi) +O(1/\sqrt
p )\right).
\end{eqnarray}
\end{lemma4}
{\it {Proof}}\,\,\, That the function is even and  entire follows
directly from the definition
\begin{equation}
\widehat{g}(p)=\int_{-1}^1\exp(ips)\cdot\exp\Big(\frac{-1}{1-s^2}\Big)ds
=\int_{-1}^1\cos(ps)\cdot\exp\Big(\frac{-1}{1-s^2}\Big)ds.
\end{equation}
The asymptotics follows from  Laplace saddle point method.
 Let
\begin{equation}
h(s,p)=-\frac{1}{1-s^2}+isp\,,
\end{equation}
denote the logarithm of the integrand. The integral is estimated
by Gaussian integrals near the appropriate extrema of $h$. The
extremum at $s_0$ contributes
\begin{equation}\label{gauss}
\sqrt{\frac{2\pi}{-h''(s_0,p)}}\cdot\exp h(s_0,p),
\end{equation}
and prime denotes a partial derivative with respect to $s$.

The extrema are the  solutions of the quartic equation
\begin{equation}
h'(s,p)=-\frac{2s}{(1-s^2)^2}+ip=0\,.
\end{equation}
 When $p$ is large two of the solutions
are close to $1$ and the other two are close to $-1$. To leading
order in $p$ the two extrema near $1$ are
\begin{eqnarray}
s_0(p)=(1\pm\delta(p)),\quad\delta(p)=
\frac{1}{\sqrt{2ip}}+O\Big(\frac{1} {p^{3/2}}\Big)\,,
\end{eqnarray}
where
\begin{equation}
h(1\pm\delta,p)= ip \pm (1+i)\sqrt{p}  -\frac{1}{4} +O(\delta).
\end{equation}
It is now clear that the right way to deform the contour is so
that it goes through $1-\delta$ and the  saddle point at
$1+\delta$ should be avoided. Now,
 to leading order in $p$
the second derivative there is
\begin{eqnarray}
h''(s_0,p)&=& -\frac{8s^2}{(1-s^2)^3}+O(p)\\ &=&-
(2ip)^{3/2}\left(1+O(\delta)\right)\nonumber \\ &=& - (2p)^{3/2}
\exp\left(\frac{3\pi i}{4}\right)\cdot
\left(1+O(\delta)\right)\nonumber
\end{eqnarray}
Substituting in Eq.~(\ref{gauss}) one finds the contribution from
the saddle points near $1$ to be
\begin{equation}
\frac{\sqrt{\pi}}{(2e)^{1/4}}\,\cdot\frac{\exp(- \sqrt
p)}{p^{3/4}}\cdot \exp \big( i(p-\sqrt p -3\pi/8 )\big)\,
\Big(1+O(1/\sqrt p )\Big).
\end{equation}
Using the fact that $\hat g(p)$ is real valued for real $p$, the
saddle near $-1$ must give the complex conjugate of this, and the
result follows. \Box
\end{appendix}

\end{document}